\documentclass{article}
\usepackage{amsmath,amssymb,epsfig}
\usepackage[tight]{subfigure}

\def\A{{\cal A}}

\def\reals{\mathbb{R}}

\def\comp{\raise 1pt \hbox{$\scriptstyle\circ$}}

\def\argmin{\mathop{\rm argmin}\limits}

\def\upto{{\raise 1pt \hbox{$\scriptstyle \,\nearrow\,$}}}
\def\downto{{\raise 1pt \hbox{$\scriptstyle \,\searrow\,$}}}

\newtheorem{model}{Model}

\begin{document}

\title{Reduced form modeling of limit order markets}

\author{Pekka Malo\footnote{Department of Business Technology, Aalto University, {\tt pekka.malo@hse.fi}}\and Teemu Pennanen\footnote{Department of Mathematics and Systems Analysis, Aalto University, {\tt teemu.pennanen@tkk.fi}, corresponding author}}

\maketitle

\begin{abstract}
This paper proposes a parametric approach for stochastic modeling of limit order markets. The models are obtained by augmenting classical perfectly liquid market models by few additional risk factors that describe liquidity properties of the order book. The resulting models are easy to calibrate and to analyze using standard techniques for multivariate stochastic processes. Despite their simplicity, the models are able to capture several properties that have been found in microstructural analysis of limit order markets. Calibration of a continuous-time three-factor model to Copenhagen Stock Exchange data exhibits e.g.\ mean reversion in liquidity as well as the so called crowding out effect which influences subsequent mid-price moves. Our dynamic models are well suited also for analyzing market resiliency after liquidity shocks.
\end{abstract}

\section{Introduction}

Most modern stock exchanges are based on the continuous double auction mechanism where outstanding limit orders are organized in the limit order book. If the size of a market order exceeds the quantity available at the best price, the trader has to climb up the book and accept worse marginal prices to get his order filled. In other words, marginal prices of market orders are increasing functions of the order size. This is in contrast with classical market models where marginal prices are stochastic processes independent of order size. 

This paper presents simple extensions to classical perfectly liquid market models in order to describe liquidity aspects of limit order markets. The models are obtained by introducing few additional risk factors that describe the liquidity properties of the order book. This allows for compact descriptions of both immediate liquidity costs of market orders as well as resiliency properties of the order book. Our models can incorporate a wide variety of existing models for market prices while leaving room for modeling liquidity aspects of the market. Classical perfectly liquid market models are obtained as special cases when illiquidity effects vanish.

There exists an extensive literature on microstructural models where arriving orders are related to various features of the limit order book; see e.g.\ Glosten~\cite{glo94}, Biais et al.~\cite{bhs95}, Parlour~\cite{par98}, Sand{\aa}s~\cite{san1}, Luckock~\cite{luc3}, Smith et al.~\cite{sfg3}, Ranaldo~\cite{ran4}, Foucalt et al.~\cite{fkk5}, Biais, Glosten and Spatt~\cite{bgs5}, Bouchaud et al.~\cite{bfl9}, Rosu~\cite{ros9} and their references. Such detailed studies provide valuable insights to the formation of the order book but the employed models may be cumbersome in describing the statistical behavior of the book over time. Recently, Cont et al.~\cite{cst8} introduced a more tractable model where orders arrive at discrete price levels according to independent Poisson processes with a rate depending on the distance to the best quotes. Their model gives a stochastic description of the complete order book which is essential e.g.\ when studying the costs of trading strategies. Several empirically observed phenomena were reproduced by the model of \cite{cst8} in numerical calculations employing Laplace transform techniques. 

The present paper proposes a further simplification by describing order book dynamics through a finite number of risk factors without explicit account of the order flow. One of the factors is the market price while the others describe liquidity properties of the order book. This reduction provides a natural link between classical perfectly liquid market models and microstructural models of limit order markets. With appropriate parameterization, essential features of the order book can be captured already with low-dimensional models. The dimensionality becomes an important issue, for example, in numerical optimization of trading strategies; see e.g.\ Bertsimas and Lo~\cite{bl98} or Koivu and Pennanen~\cite{kp10}.

Another advantage of our approach is its simplicity. The resulting models are easy to calibrate and to analyze using standard techniques for multivariate stochastic processes. This is illustrated by calibrating continuous-time order book models to limit order data from Copenhagen stock exchange. We model the market price and two liquidity factors by a three-dimensional stochastic differential equation which is calibrated to historical observations of order books. Despite its simple structure, the model is able to reproduce some well-known properties of limit order markets. In particular, the calibrated models are mean reverting in liquidity and they exhibit the so called ``crowding out'' effect, which means that high liquidity on one side of the book encourages more aggressive limit orders and quote improvements on that side. In our reduced form model, this shows up as increases in the mid-price when the ask-side liquidity, in terms of the liquidity factors, is high compared to the liquidity on the bid-side and vice versa. Both effects are supported by extensive theoretical and empirical studies of limit order markets; see e.g.\ Biais et al.~\cite{bhs95}, Foucault, Kadan and Kandel~\cite{fkk5}, Parlour~\cite{par98}, Griffiths et al.~\cite{gstw}, Ranaldo~\cite{ran4}, Bouchaud et al.~\cite{bfl9}.

Our models are close to those developed in {\c{C}}etin, Jarrow and Protter~\cite{cjp4} and {\c{C}}etin, Jarrow, Protter and Warachka~\cite{cjpw6}. The main difference is that our models retain the monotonicity of marginal prices of market orders which implies that their total costs are convex. Convexity is an essential feature of limit order markets and it has many important implications for risk management and pricing and hedging of financial instruments; see e.g.\ Delbaen and Schachermayer~\cite{ds6} or F\"ollmer and Schied~\cite{fs4} for classical perfectly liquid models and Jouini and Kallal~\cite{jk95a} or Kabanov~\cite{kab99} for models with proportional transactions costs. Our models can be seen as generalizations of the model of {\c{C}}etin and Rogers~\cite{cr7} who studied utility maximization under liquidity costs given in terms of a fixed convex function of the order size. Basic arbitrage and pricing theory for general convex cost functions (that cover, in particular, piecewise linear functions associated with limit order markets) has been developed in Pennanen~\cite{pen8,pen8b}. Besides theoretical aspects, convexity is essential also in optimization of trading strategies; see e.g.\ Edirisinghe, Naik and Uppal~\cite{enu93}, Bertsimas and Lo~\cite{bl98}, Almgren and Chriss~\cite{ac}, Huberman and Stanzl~\cite{hs5}, Alfonsi, Schied and Fruth~\cite{asf10} or Koivu and Pennanen~\cite{kp10}.

The rest of this paper is organized as follows. Section~\ref{sec:lob} gives a short review of the order book structure and a quantification of liquidity costs in monetary terms. This is used in Section~\ref{sec:model} to describe liquidity costs by simple parametric models driven by few liquidity factors. The stochastic order book models are then obtained by specifying stochastic models for the behavior of the factors. Section~\ref{sec:emp} presents an empirical study using data from Copenhagen stock exchange and Section~\ref{sec:conc} concludes.

\section{Liquidity costs of market orders}\label{sec:lob}

We begin with a quick review of order book structure in order to introduce the main concepts, the data and the notation. The limit order book (LOB) maintains a record of all submitted limit orders that have not been cancelled or met by a market order. Table~\ref{tab:lob1} displays a part of a limit order book of one particular stock at the Copenhagen stock exchange at a given time instant. When submitting a buy market order, only a finite number of shares can be bought at the best ask price and when buying more, one gets the second lowest price and so on. For example, exactly 20800 shares could have been bought at the best ask price of 239 in the LOB of Table~\ref{tab:lob1}.

\begin{figure}
\begin{center}
\epsfig{file=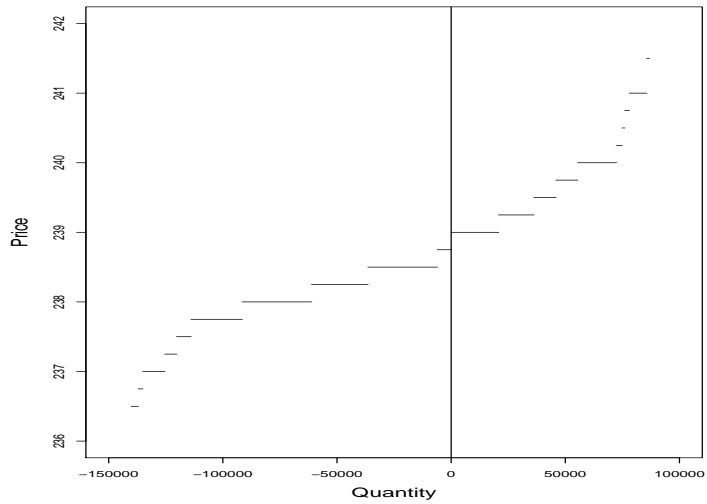,height=0.6\linewidth,width=0.8\linewidth, angle=0}
\end{center}
\caption{Marginal price curve corresponding to Table~\ref{tab:lob1}. The horizontal axis gives the cumulative depth of the book measured in the number of shares. Negative order quantity corresponds to a sale. The price unit is 1 DKK.\label{fig:lob}}
\end{figure}

The {\em marginal price} of a buy market order is a positive, nondecreasing, piecewise constant function of the number of shares bought. For sell market orders, the situation is similar and the {\em marginal revenue} of selling is a positive, nonincreasing, piecewise constant function of the number of shares sold. Interpreting sell orders as buy orders of negative quantity, we can, as e.g.\ in Glosten~\cite{glo94}, incorporate the bid and ask sides of the book into a single curve $s:\reals\mapsto[0,+\infty]$ that gives the marginal price of a market order of arbitrary sign and quantity. Since the ask price is always strictly greater than the bid price, $s$ is monotone and positive. Figure~\ref{fig:lob} plots the marginal price curve associated with the LOB in Table~\ref{tab:lob1}.


\begin{table}
\caption{Example limit order book. The table presents a snapshot of TDC A/S limit order book observed at 12JAN2005:13:58:19.430. The book is obtained from Copenhagen Stock Exchange order flow data by implementing the rules of SAXESS trading protocol.}\label{tab:lob1}
{\footnotesize
\begin{center}
\begin{tabular}{c c | c c}
\hline
\multicolumn{2}{c}{Bid}  &  \multicolumn{2}{c}{Ask} \\
    \hline
Price   &   Quantity   &   Price   &   Quantity   \\
238.75  &   140 &   239 &   3700   \\
238.75  &   600 &   239 &   1000   \\
238.75  &   3300    &   239 &   5000   \\
238.75  &   2000    &   239 &   1000   \\
238.5   &   10000   &   239 &   1000   \\
238.5   &   3900    &   239 &   2500   \\
238.5   &   15000   &   239 &   6600   \\
238.5   &   1500    &   239.25  &   10000   \\
238.25  &   10000   &   239.25  &   2500    \\
238.25  &   1000    &   239.25  &   3000    \\
238.25  &   3500    &   239.5   &   600    \\
238.25  &   10000   &   239.5   &   5000   \\
238.25  &   200   &   239.5   &   800   \\
\vdots  &  \vdots     &   \vdots    &  \vdots\\
\hline
\end{tabular}
\end{center}
}
\end{table}


The monotonicity of the marginal price curve $s$ is essential when studying trading costs. The positivity and monotonicity of $s$ imply that the {\em total cost} 
\begin{equation*}
S(x):=\int_0^xs(z)dz
\end{equation*}
of a market order of $x$ shares is a nondecreasing convex function that vanishes at $x=0$. A negative $x$ incurs a negative cost which just means that sales yield revenue. Since $s$ is piecewise constant, $S$ is piecewise linear. In a perfectly liquid market, the marginal price $s$ would be constant and the cost function $S$ would be linear. Consequences of nonlinearities on arbitrage theory and on pricing and hedging have been studied in a number of papers. Jouini and Kallal~\cite{jk95a} studied the effect of a bid-ask spread with infinite supply at the best quotes. {\c{C}}etin, Jarrow and Protter~\cite{cjp4} studied markets with nonlinear illiquidity effects in terms of the ``supply curve'', which in our notation corresponds to the function $x\mapsto S(x)/x$. {\c{C}}etin and Rogers~\cite{cr7} studied portfolio optimization in markets with differentiable convex cost functions. General convex cost functions that cover, in particular, piecewise linear functions associated with limit order markets have been studied in Pennanen~\cite{pen8,pen8b}. There the aim was to study the effects of nonlinearities on arbitrage and superhedging theories.

Perfectly liquid market models are often taken as approximations of real markets where the marginal price curve is replaced by a horizontal line passing through the {\em mid-price} (``market price'')
\begin{equation*}
\bar s := \frac{s^b + s^a}{2},
\end{equation*}
where $s^b$ and $s^a$ are the best bid and best ask price, respectively. In terms of the cost of market orders,
\begin{equation*}
s^b := \lim_{x\upto 0}\frac{S(x)}{x}\quad\text{and}\quad s^a := \lim_{x\downto 0}\frac{S(x)}{x}.
\end{equation*}
Modeling of $\bar s$ is a classical subject in financial econometrics. In this paper, we focus on modeling the nonconstancy of the marginal price curve around the mid-price $\bar s$.

From practical point of view, it is more interesting to describe liquidity in terms of monetary units rather than in numbers of shares. For example, one would not expect the shape of an order book to remain the same after a stock split; see also \cite[Section~4.4]{cjpw6}. While market values of shares are usually regarded as upward drifting processes it might be that illiquidity effects are of more stationary nature in the long run when measured in monetary units. A simple monetary measure of illiquidity is given by the function
\begin{equation*}
r(h) := \ln\frac{s(h/\bar s)}{\bar s} = \ln s(h/\bar s) - \ln\bar s
\end{equation*}
which gives the {\em percentual} change in the marginal price relative to the mid-price, as a function of the {\em mark-to-market value} $h=\bar sx$ of a market order of $x$ shares. Taking logarithms of the whole marginal price curve is analogous to the common practice of modeling market price dynamics by describing the behavior of its logarithm.

\begin{figure}
\begin{center}
\epsfig{file=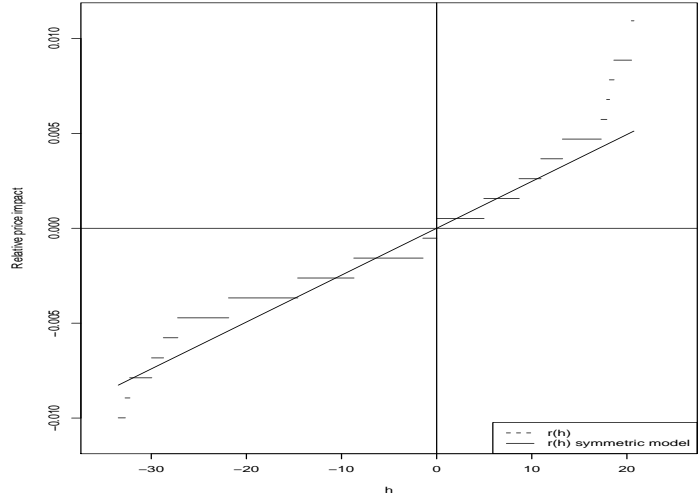,height=0.6\linewidth,width=0.8\linewidth, angle=0}
\caption{The price impact curve $r$ (step function) corresponding to Table~\ref{tab:lob1} together with its linear approximation. The horizontal axis gives the mark-to-market value $h$ of the order quantity. The unit of $h$ is DKK 1 Million.\label{fig:r1}}
\end{center}
\end{figure}

We will call $r$ the {\em relative price impact curve}. It describes the {\em temporary} price impact that a market order would have on the best quotes (bid or the ask, depending on the sign of the order). It is always nondecreasing and passes through the origin. In perfectly liquid markets $r\equiv 0$, whereas steeper $r$ corresponds to illiquidity in execution of market orders. Several authors have studied how trades affect the expected mid-price at some point in the future; see e.g.\ Hasbrouck~\cite{has91}, Bouchaud et al.~\cite[Section~5]{bfl9} and the references therein. Such dependencies are often called ``price impacts'' but they should not be confused with the relative price impact curve $r$ which describes the directly observable instantaneous impacts of market orders. The relative price impact curve $r$ corresponding to the LOB given in Table~\ref{tab:lob1} is drawn as the step function in Figure~\ref{fig:r1}.

The marginal price curve $s$ and the cost function $S$ can be expressed in terms of the mid-price $\bar s$ and the relative price impact curve $r$ as
\begin{equation*}
s(x) = \bar se^{r(\bar s x)}
\end{equation*}
and
\begin{equation*}
S(x) = \int_0^x\bar se^{r(\bar s z)}dz,
\end{equation*}
respectively. Making the change of variables $x=h/\bar s$, we get the representation
\begin{equation}\label{phi}
S(x) = \varphi(\bar s x),
\end{equation}
where
\begin{equation*}
\varphi(h) := \int_0^he^{r(z)}dz.
\end{equation*}
The function $\varphi$ is convex, nondecreasing and vanishes at the origin. It satisfies $\varphi(h)\ge h$, where equality holds in perfectly liquid markets. Indeed, since $r$ is a nondecreasing function passing through the origin, we have $e^{r(h)}\ge 1$ for $h\ge 0$ and $e^{r(h)}\le 1$ for $h\le 0$. Integrating from $0$ to $h$ gives $\varphi(h)\ge h$. Equality holds exactly when $r\equiv 0$.



The function $\varphi$ gives the cost of a market order whose {\em mark-to-market value} is $h$. The decomposition of $S$ in \eqref{phi} supports the common practice of regarding market risk and liquidity risk as separate sources of financial risks. Whereas unpredictable variations in the mid-price $\bar s$ are often interpreted as {\em market risk}, variations in $\varphi$ (or, equivalently, in $r$) may be interpreted as {\em liquidity risk}. In the remainder of this paper, we focus on modeling the liquidity cost function $\varphi$ and its relation to the mid-price $\bar s$.

\section{Reduced form LOB models}\label{sec:model}


In classical perfectly liquid market models, the mid-price $\bar s$ is a positive stochastic process and it is assumed that marginal prices of market orders are independent of order size. This corresponds to modeling the limit order book by a horizontal line passing through the mid-price $\bar s$ so that $r\equiv 0$. In this section we propose simple extensions where the shape of the limit order book is described by modeling the relative price impact curve $r$ as a simple parametric function of the order size. The parameters may be interpreted as risk factors that describe liquidity properties of the market. A stochastic order book model is then obtained by describing the mid-price and the liquidity factors as a multivariate stochastic process.

A minimal condition for a sensible market model is that the marginal price curve $s$ is positive and nondecreasing, or equivalently, that the cost function $S$ is increasing and convex. This can be achieved by expressing the cost function in terms of the relative price impact curve as in \eqref{phi} and modeling the price impact curve $r$ as a nondecreasing curve passing through the origin\footnote{This is parallel to the common practice of modeling the term structure of interest rates by describing the behavior of the forward curve. The positivity of the forward curve guarantees that the zero curve is decreasing as a function of maturity and that it lies between zero and one. 
}.
A simple model that looks beyond the bid and ask prices is obtained with  
\begin{equation}\label{e:lin}
r(h) = \beta h,
\end{equation}
where $\beta>0$ describes the overall degree of illiquidity in the book. Larger the value of $\beta$, less liquid the market. This is reminiscent of the linear specification of the ``price impact function'' in Sand{\aa}s~\cite{san1}. However, the price impact function in Sand{\aa}s~\cite{san1} describes the dependence of the ``fundamental value'' of the stock (one period ahead) on the order quantity while the relative price impact curve $r$ describes the shape of the limit order book which is directly observable. 

The solid line in Figure~\ref{fig:r1} corresponds to the average slope
\begin{equation*}
\beta :=\argmin_{\beta\in\reals} \int_{h^-}^{h^+}|r(h)-\beta h|^2dh
\end{equation*}
of the relative price impact curve of the LOB in Table~\ref{tab:lob1}. Here $h^+>0$ and $h^-<0$ are finite cut-off parameters that limit the focus on the best 10 prices on each side of the book. The linear model \eqref{e:lin} corresponds to
\begin{equation*}
\varphi(h) = \int_0^he^{r(z)}dz = \frac{e^{\beta h}-1}{\beta}
\end{equation*}
and thus, to the cost function
\begin{equation*}
S(x) = \frac{e^{\beta\bar s x}-1}{\beta}.
\end{equation*}
This is close to the exponential model studied in {\c{C}}etin and Rogers~\cite{cr7}. In their model $S(x)=\bar s(e^{\alpha x}-1)/\alpha$ which describes liquidity effects in terms of units of shares and implies, in particular, that illiquidity effects diminish when the market price $\bar s$ increases and vice versa.

\begin{figure}
\begin{center}
\epsfig{file=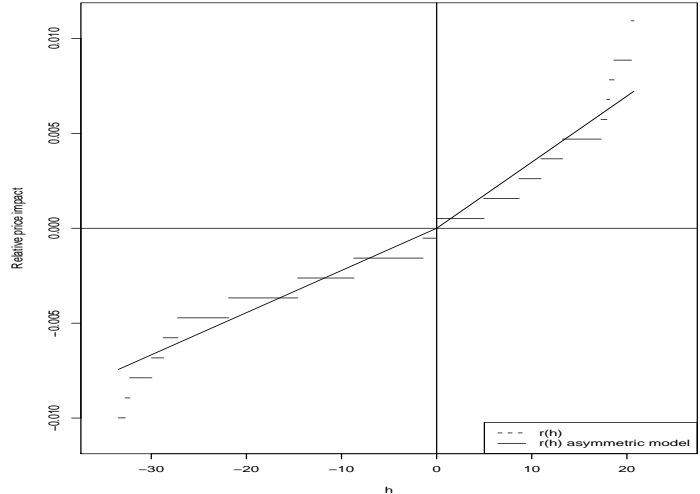,height=0.6\linewidth,width=0.8\linewidth, angle=0}
\caption{The price impact curve $r$ and its two sided linear approximation. The horizontal axis gives the mark-to-market value $h$ of the order quantity. The unit of $h$ is DKK 1 Million. \label{fig:r2}}
\end{center}
\end{figure}

In some situations, the bid and ask sides of a limit order book exhibit different degrees of illiquidity and the numbers
\begin{align*}
\beta^+&:=\argmin_{\beta\in\reals} \int_0^{h^+}|r(h)-\beta h|^2dh,\\
\beta^-&:=\argmin_{\beta\in\reals} \int_{h^-}^0|r(h)-\beta h|^2dh,
\end{align*}
may be significantly different. Differences in liquidity on the two sides of the book is known to be significant in explaining e.g.\ mid-price movements; see Parlour~\cite{par98}, Griffiths et al.~\cite{gstw}, Ranaldo~\cite{ran4} and Cont et al.~\cite{cst8}. We will model the relative price impact curve by the piecewise linear function
\begin{equation}\label{eq:r2}
r(h) = 
\begin{cases}
\beta^+h & \text{for $h\ge 0$},\\
\beta^-h & \text{for $h\le 0$}.
\end{cases}
\end{equation}
Figure~\ref{fig:r2} plots the corresponding approximation of the LOB in Table~\ref{tab:lob1}. This corresponds to the cost function
\begin{equation*}
S(x) = 
\begin{cases}
\frac{e^{\beta^+\bar sx}-1}{\beta^+} & \text{for $x\ge 0$},\\
\frac{e^{\beta^-\bar sx}-1}{\beta^-} & \text{for $x\le 0$}
\end{cases}
\end{equation*}
This could be extended in an obvious way to cover nonzero bid-ask spreads or proportional transaction costs as e.g.~in Jouini and Kallal~\cite{jk95a} (who assumed infinite supply at the best quotes). This would correspond to a discontinuity of $r$ at the origin. The bid-ask spread could be taken as an additional liquidity factor in the model.

Given a parametric description of the order book, we obtain stochastic market models with illiquidity effects by modeling the parameters as a multivariate stochastic process. If the relative price impact curve is parameterized by a vector $\theta\in\reals^d$, then the marginal price curve is given by
\begin{equation*}
s(x) = \bar se^{r(\bar sx,\theta)}.
\end{equation*}
The corresponding cost function can be expressed as in \eqref{phi} as
\begin{equation*}
S(x) = \int_0^{\bar s x}e^{r(z,\theta)}dz.
\end{equation*}
We can employ the wide variety of existing stochastic models for the mid-price $\bar s$ and augment them with stochastic descriptions of $\theta$. The corresponding cost functions define a convex cost process in the sense of \cite{pen8} and are thus amenable to analytical techniques similar to those applied to classical perfectly liquid market models; see~\cite{pen8} and \cite{pen8b}. The concrete interpretations of the parameters are useful when specifying the stochastics. 

\begin{model}
If the relative price impact curve is parameterized by \eqref{eq:r2}, then the market model is driven by the three-dimensional process $(\bar s,\beta^-,\beta^+)$. A simple stochastic specification is to assume a univariate process for the mid-price $\bar s$ and that the two dimensional vector $\theta=(\ln\beta^-,\ln\beta^+)$ follows a two-dimensional Ornstein-Uhlenbeck process
\begin{equation*}
d\theta=\Gamma(\mu-\theta)dt + \Sigma dW_t,
\end{equation*}
where $\Theta\in\reals^{2\times 2}$, $\mu\in\reals^2$ and $\Sigma\in\reals^{2\times 2}$ are parameters of the model and $W$ is e.g.\ a two-dimensional Brownian motion. 
\end{model}

The following allows for modeling dependencies between the mid-price and the liquidity factors.

\begin{model}\label{ex:lin}
Assume that the three-dimensional process $\xi=(\ln\bar s,\ln\beta^-,\ln\beta^+)$ satisfies the linear stochastic differential equation
\begin{equation}\label{eq:lin}
d\xi = (A\xi+a)dt + \Sigma dW_t,
\end{equation}
where $A\in\reals^{3\times 3}$ and $a\in\reals^3$. In particular, the mid-price drift may depend on the shape of the order book through the liquidity factors $\beta^-$ and $\beta^+$; see Section~\ref{sec:emp}. The solution of \eqref{eq:lin} can be written as
\begin{equation*}
\xi_t = e^{tA}\xi_0 + \int_0^te^{(t-s)A}ads + \int_0^te^{(t-s)A}\Sigma dW_s.
\end{equation*}
It seems reasonable to assume that the drift is stationary in the sense that there is a $\delta\in\reals^3$ such that
\begin{equation*}
\lim_{t\to\infty}E[A\xi_t+a] = \delta.
\end{equation*}
In particular, we would expect that the liquidity factors are stationary so that the second and the third component of $\delta$ are zero. Unless the mid-price is also stationary, we must then have that the first column of $A$ is zero since otherwise the expectation could not exist. The average values of the liquidity factors are then determined by
\begin{equation*}
\begin{bmatrix}
A_{22} & A_{23}\\
A_{32} & A_{33}
\end{bmatrix}
\begin{bmatrix}
E \beta^-\\
E \beta^+
\end{bmatrix}
+\begin{bmatrix}
a_2\\
a_3
\end{bmatrix}
=
\begin{bmatrix}
0\\
0
\end{bmatrix}
\end{equation*}
and the average mid-price drift becomes
\begin{equation*}
\delta_1 = A_{12}E \beta^- + A_{13}E \beta^+ + a_1.
\end{equation*}
\end{model}

\section{Empirical study}\label{sec:emp}

As an illustration of the modeling approach described above, we will calibrate the three-dimensional Model~\ref{ex:lin} to order book data from Copenhagen Stock Exchange. We first adjust the liquidity factors for intra-day patterns that are found to explain a considerable part of the variations. The three-dimensional stochastic differential equation is then calibrated to discrete observations of the mid-price and the deseasonalized liquidity factors. The resulting model exhibits certain well-known features of limit order markets.

\subsection{Data}

We use a data set from Copenhagen Stock Exchange (CSE) that contains the buy and sell orders and transactions of all traded stocks for the period January 2005 to March 2005.\footnote{We thank Nikolaj Munck at OMX market research in Stockholm for providing the data set and assisting with issues involved in reconstruction of the limit order books.} Each order record carries information about the date and time of submission, the order type, the quantity and the price. Trading on CSE takes place in the common trading system SAXESS, which implements a continuous double auction mechanism. In this study we focus only on the trading lot market and have removed all odd lots from the dataset. For a detailed description of the trading conditions and characteristics of the system we refer to NOREX Member Rules book\footnote{See http://www.nasdaqomx.com/listingcenter/nordicmarket/rulesandregulations/copenhagen for the latest copy of NOREX Member Rules. The results in this paper are based on the rules effective over the data period January--March 2005}.

We use the event history to construct the order books and the corresponding mid-price $\bar s$ and the bid and ask side liquidity factors $\beta^-$ and $\beta^+$ as described in Section~\ref{sec:lob}. The computations are carried out using SAS and R computing environments. Starting from the initial state of the order book at the beginning of the trading day, we update the book at every new event by implementing the rules of SAXESS trading protocol. All new orders, deletions and updates of earlier orders and trade executions are accounted for as they arrive. We use the order's direction, price and position to determine its rank in the book according to price/time priority. When constructing the books, we do not include hidden volume (contained e.g.\ in iceberg orders) in our computations to ensure that the books correspond to the information available to traders over time. Also, to eliminate the effects pre- and post-trading sessions, we exclude the first and the last trading hours from the data.

\subsection{Intra-day patterns in liquidity}\label{sec:idp}

\begin{figure}
\begin{center}
\epsfig{file=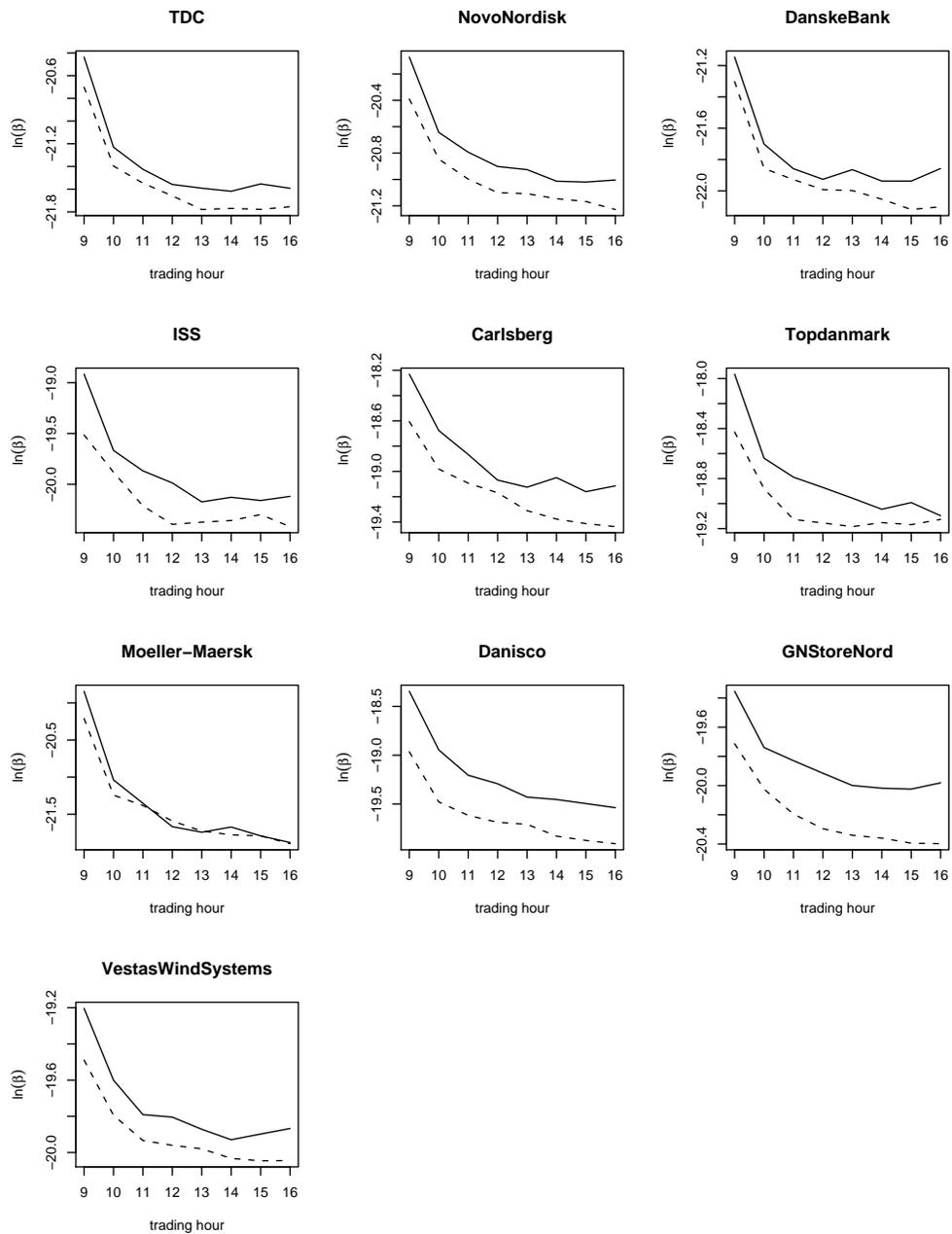,width=1.1\linewidth, angle=0}
\caption{Intra-day patterns in the liquidity factors. The two lines in each figure give the estimated hourly dummies for the liquidity factors. The solid line corresponds to the bid-side.}\label{fig:diurnal}
\end{center}
\end{figure}

As discussed e.g.\ in Engle and Russell~\cite{er9}, the frequency of trades, volume, and spreads all typically exhibit consistent patterns over the course of the day. Therefore, before calibrating our model, we account for intra-day liquidity patterns by regressing $\ln\beta_t^+$ and $\ln\beta_t^-$ on trading hour dummies. Since the liquidity factors are all strictly positive we have taken logarithms before the linear regressions. The results for 10 stocks are plotted in Figure~\ref{fig:diurnal}. All stocks exhibit decreasing values of $\beta^+$ and $\beta^-$ over the trading day. This means that the liquidity of the order book tends to increase as more orders are accumulated over the day. This is consistent with earlier studies of intra-day patterns in order books. For example, Kempf and Mayston~\cite[Section~2]{kem8} observed that the spread decreases and the depth of the book increases over the trading day.

Another notable pattern is that the bid side seems steeper on average than the ask side. This may reflect the fact that traders on the buy side are exposed to market crashes which are often perceived more likely than sudden upward moves of similar magnitude. On the other hand, given the relatively short time period, this phenomenon may also be related to market conditions and may be reversed e.g.\ in bull markets.

\begin{figure}
\begin{center}
\epsfig{file=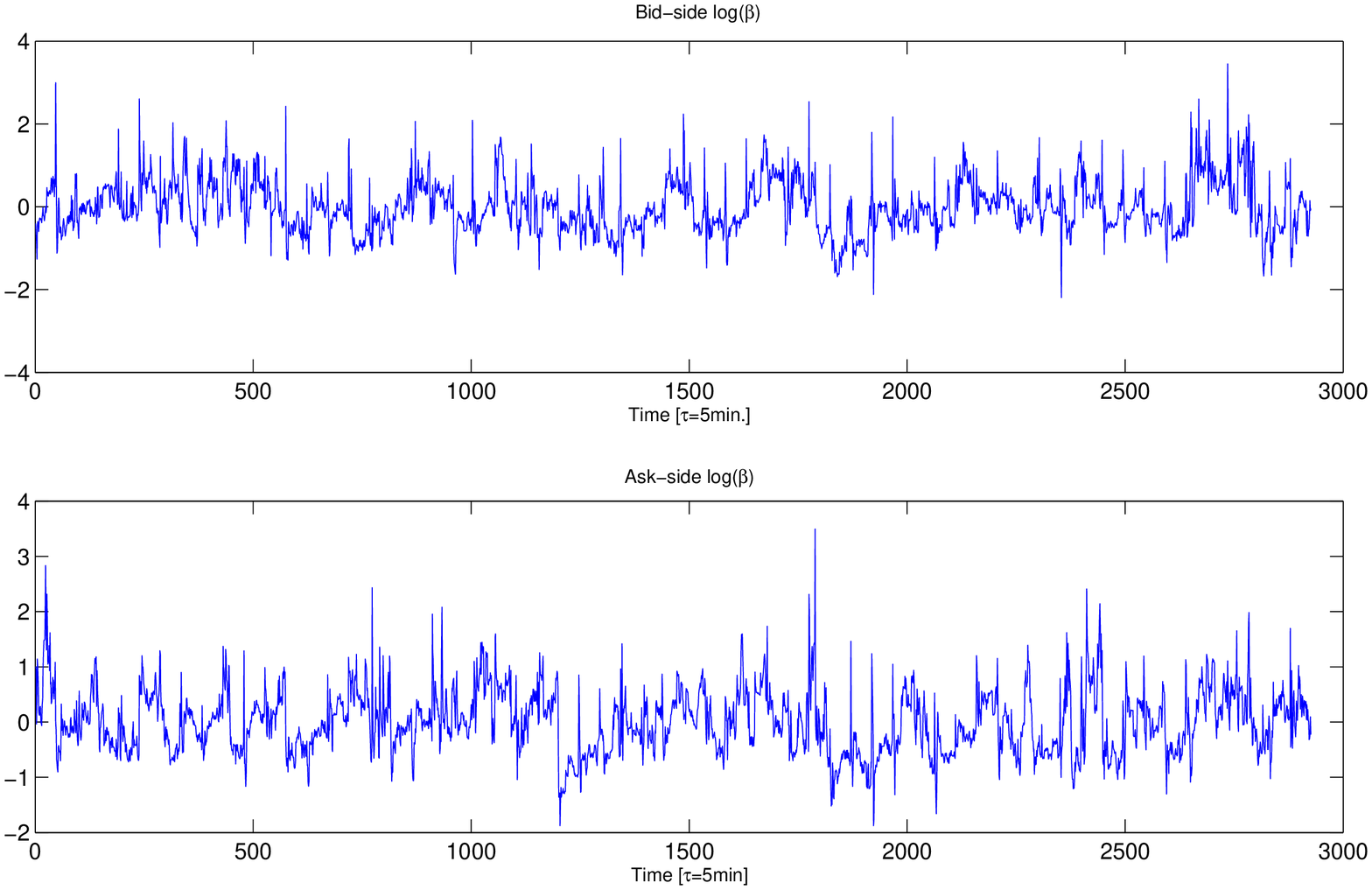,width=\linewidth, angle=0}
\epsfig{file=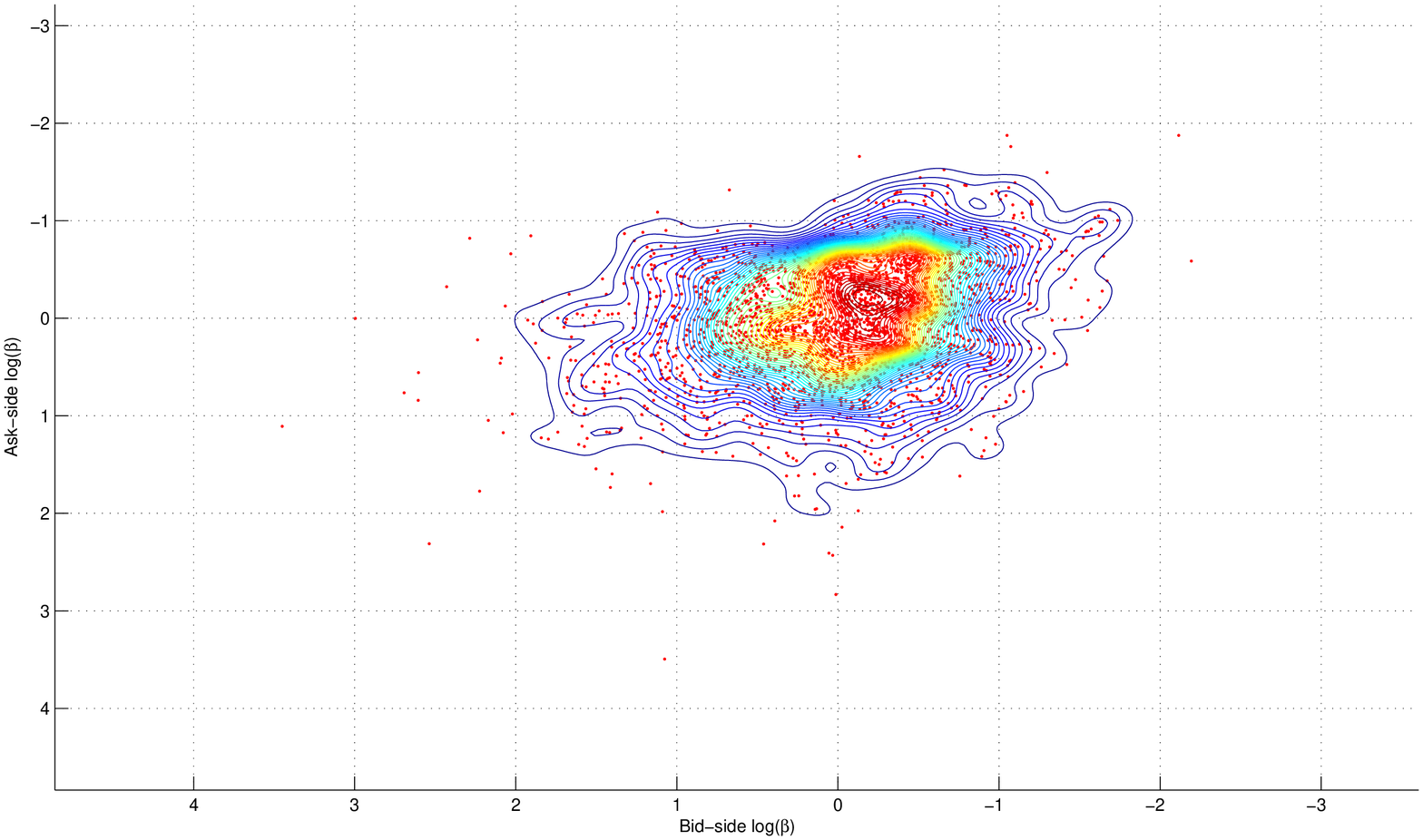,width=\linewidth,height=0.6\linewidth, angle=0}
\caption{The time series plots of the deseasonalized bid- and ask-side liquidity factors, $\ln \beta^-$ and $\ln \beta^+$, for TDC during January 2005 - March 2005. The lower picture gives the contour plots of the corresponding bivariate kernel-density estimate. The time interval is 10 minutes.}\label{fig:tdc}
\end{center}
\end{figure}

\begin{figure}
\begin{center}
TDC
\epsfig{file=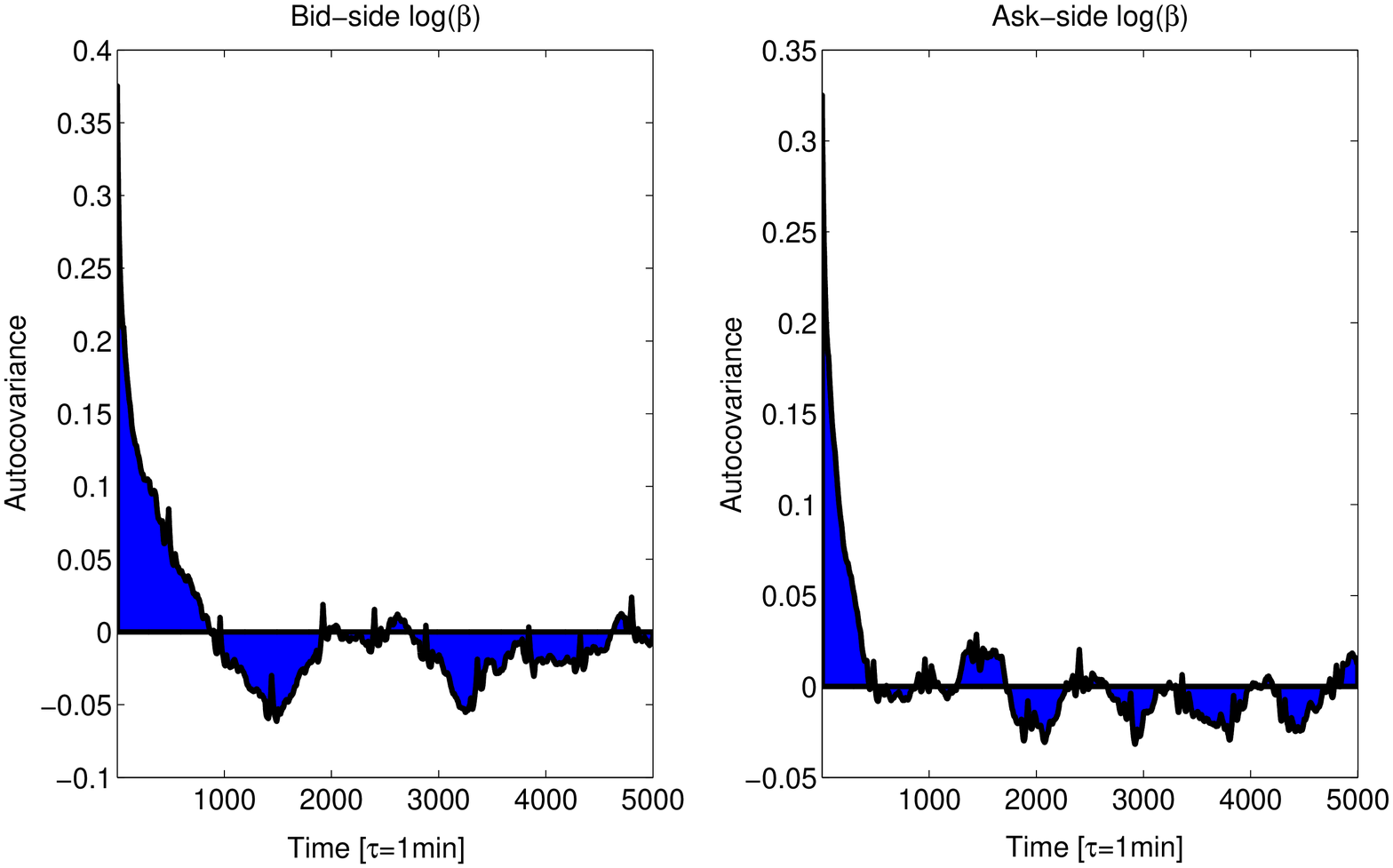,height=0.3\linewidth,width=\linewidth, angle=0}
Moeller-Maersk
\epsfig{file=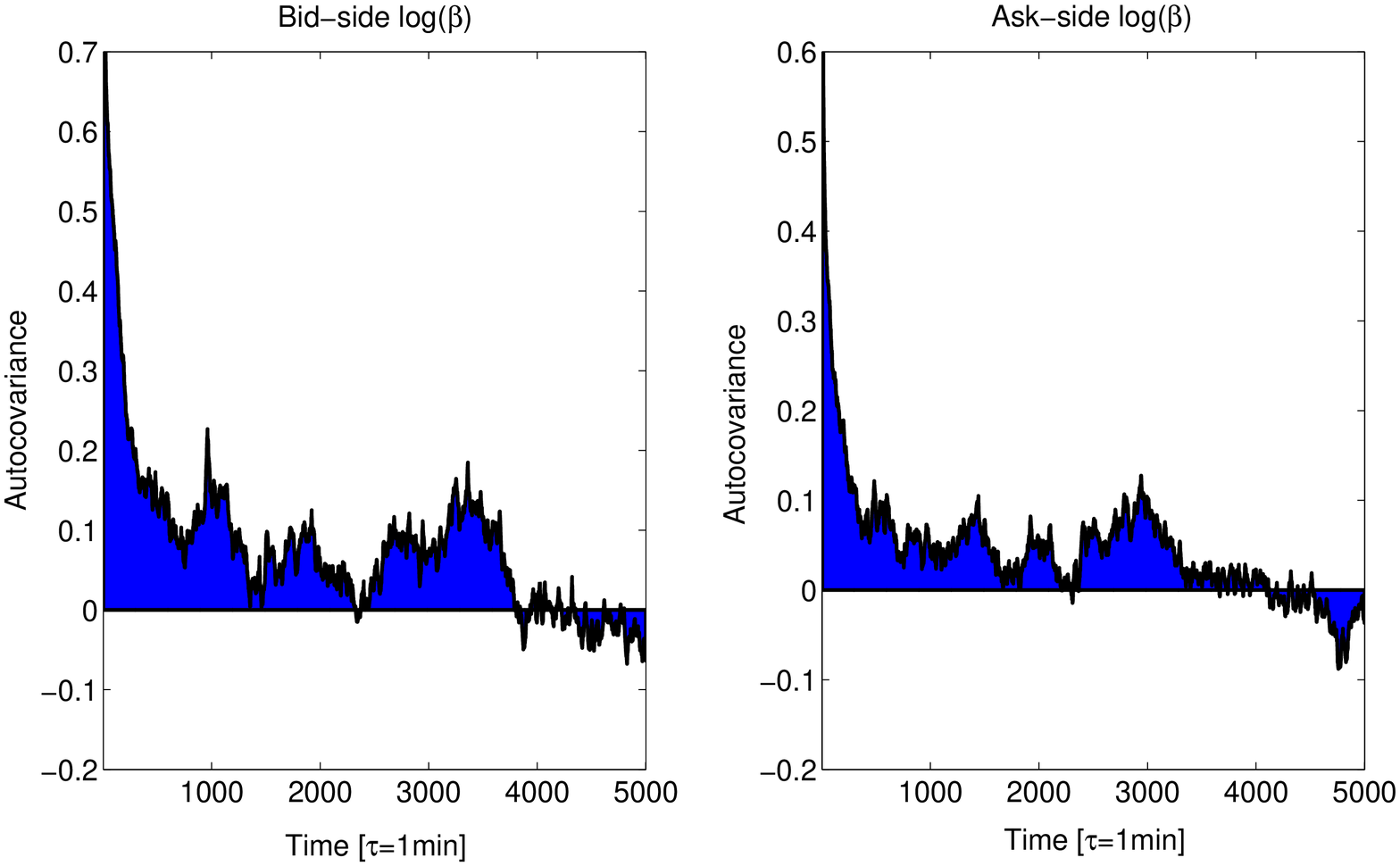,height=0.3\linewidth,width=\linewidth, angle=0}
\caption{The autocovariance functions of $\ln\beta^-$ and $\ln\beta^+$ for TDC and Moeller-Maersk during January 2005 - March 2005. The time interval is 1 minute.}\label{fig:autocovs}
\end{center}
\end{figure}

From now on, the symbols $\beta^+$ and $\beta^-$ will denote the deseasonalized variables. Figure~\ref{fig:tdc} plots the evolution of the logarithms of $\beta^+$ and $\beta^-$ together with contour plots of the kernel density estimate of their joint probability density for TDC's limit order book. The series is obtained with 10 minute sampling interval. The liquidity factors present significant variation over time. The findings are similar for the rest of the stocks. Figure~\ref{fig:autocovs} plots the autocovariance functions of the liquidity factors for TDC and Moeller-Maersk. The dynamic behavior of the factors are characterized by strong positive autocorrelations which decay exponentially.

\subsection{Calibration of a linear model}

Consider Model~\ref{ex:lin} where the order book is driven by the three dimensional vector $\xi=(\ln\bar s, \ln\beta^-,\ln\beta^+)$. For $l>0$, the conditional distribution of $\xi_l$ given $\xi_0$ is Gaussian with mean
\begin{equation*}
\bar\xi_l = e^{lA}\xi_0 + \int_0^le^{(l-s)A}ads
\end{equation*}
and variance (by Ito isometry)
\begin{equation}\label{var}
V_l = \int_0^le^{(l-s)A}\Sigma[e^{(l-s)A}\Sigma]^Tds.
\end{equation}
In other words,
\begin{equation}\label{tsm}
\xi_l = B_l\xi_0 + b_l + \varepsilon,
\end{equation}
where $B_l=e^{lA}$, 
\begin{equation*}
b_l = \int_0^le^{(l-s)A}ads
\end{equation*}
and $\varepsilon$ is Gaussian with zero expectation and variance $V_l$. The model can be calibrated to discrete data with fixed time increments $l$ by first estimating the values of $B_l$, $b_l$ and $V_l$ and then solving for the corresponding $A$, $a$ and $\Sigma$. Indeed, \eqref{tsm} can be seen as a linear time series model whose parameters can be estimated with ordinary least squares. Once estimates of $B_l$, $b_l$ and $V_l$ are available, the matrix $A$ can be estimated with the matrix logarithm
\begin{equation*}
A=\frac{1}{l}\ln B_l.
\end{equation*}
On the other hand
\begin{equation*}
a=\A_l^{-1}b_l,
\end{equation*}
where the matrix
\begin{equation*}
\A_l = \int_0^l e^{(l-s)A}ds.
\end{equation*}
can be computed numerically by solving the initial value problem
\begin{equation*}
\begin{cases}
\A_0 = 0 \\
\frac{d}{dt}\A_t = I + A\A_t
\end{cases}
\end{equation*}
using techniques of numerical integration\footnote{In this study, we employ MATLAB routine ODE45.}. The volatility matrix $\Sigma$ can be obtained with a similar technique using the identity \eqref{var} and the estimated values of $A$ and $V_l$

We used the above procedure to fit Model~\ref{ex:lin} to order book data with the sampling interval $h$ equal to 10 minutes. Table~\ref{tab:ou} gives the results for TDC and Moeller-Maersk. The estimates of the matrix $A$ are consistent with the empirical autocovariance structures and the observed mean-reverting behavior of liquidity; see e.g.\ \cite{bhs95} or \cite{fkk5}. This shows up as the negative eigenvalues of $A$ given in Panel B of Table~\ref{tab:ou}. The negative eigenvalues imply, in particular, that the liquidity factors are mean reverting. In the words of Biais et al.~\cite[page 1657]{bhs95}, ``investors provide liquidity when it is valuable to the marketplace and consume liquidity when it is plentiful''. This can be explained e.g.\ by the so called ``stimulated refill'' effect which means that when liquidity is reduced by a market order on one side of the book new limit orders appear more frequent on that side; see e.g.\ \cite{bfl9} or \cite{bkp6}.

The eigenvalue which is close to zero corresponds to the mid-price which is usually found to be nonstationary. In our three-month dataset, even the mid-price appears stationary in some cases. It can be seen as a consequence of the mean reversion property of the mid-price which is often reported for high-frequency data of stock prices; see e.g.\ \cite{er9} or Harris and Panchapagesan~\cite{hp5}. This corresponds to the negative sign of the first element on the first row of $A$.

The signs of the second and the third element on the first row of $A$ suggest that the order book shape affects mid-price movements. In particular, the negativity of the second element means that a steep bid-side lowers the drift of the mid-price; see Model~\ref{ex:lin}. On the ask side, the effect is reversed. This corresponds to the so called ``crowding out'' effect which is supported by several studies of limit order markets; see e.g.\ Parlour~\cite{par98}, Griffiths et al.~\cite{gstw}, Ranaldo~\cite{ran4} and Cont et al.~\cite{cst8}. It means that increased supply on one side of the book results in more aggressive incoming orders on that side and less aggressive orders on the opposite side. In our reduced form model, this shows up as increased mid-price drift when the ask-side liquidity, in terms of the liquidity factors, is high compared to the liquidity on the bid-side and vice versa. The results were found to be consistent across different stocks and different sampling intervals.


\begin{table}
\caption{Parameter estimates for TDC and Moeller-Maersk. The panel A gives the parameter estimates for the linear equation $d\xi = (A\xi+a)dt + \Sigma dW_t$ in Model~\ref{ex:lin}, where $\xi=(\ln\bar{s},\ln\beta^-,\ln\beta^+)$. Panel B gives the spectral decomposition of the matrix $A$ for both stocks.}\label{tab:ou}
\vspace{2mm}
\centering
\begin{tabular}{lcc}
\multicolumn{3}{l}{Panel A: Parameters for Model~\ref{ex:lin}} \\
\hline
Parameter & TDC & Moeller-Maersk \\
\hline
\\
 $A$ &$\begin{pmatrix}
-0.0014	&	-0.0003	&	0.0004	\\
-0.5132	&	-0.2466	&	-0.0035	\\
-0.9445	&	-0.0133	&	-0.1952	\\

\end{pmatrix}$ & 
$\begin{pmatrix}
-0.0002	&	-0.0001	&	0.0001	\\
0.9247	&	-0.6417	&	0.1336	\\
-1.3643	&	0.1465	&	-0.7363	\\
\end{pmatrix}$ \\
\\
$\Sigma$ & 
$\begin{pmatrix}
0.0000	&	-0.0001	&	0.0000	\\
-0.0001	&	0.1963	&	0.0492	\\
0.0000	&	0.0492	&	0.1254	\\
\end{pmatrix}$ &
$\begin{pmatrix}
0.0000	&	0.0001	&	-0.0002	\\
0.0001	&	2.0846	&	-0.0492	\\
-0.0002	&	-0.0492	&	1.7570	\\
\end{pmatrix}$ \\
\\
$a$ & 
$\begin{pmatrix}
0.0080	\\
2.8240	\\
5.1970	\\
\end{pmatrix}$ &
$\begin{pmatrix}
0.0022	\\
-9.9966	\\
14.7467	\\
\end{pmatrix}$ \\
\\

\hline
\end{tabular}

\vspace{2mm}
\begin{tabular}{lcc}
\multicolumn{3}{l}{Panel B: Spectral decomposition of $A$}\\
\hline
 & TDC & Moeller-Maersk \\
\hline
\\
Eigenvectors &$\begin{pmatrix}
-0.1893	&	0.0007	&	-0.0023	\\
0.3857	&	0.9664	&	-0.0424	\\
0.9030	&	0.2569	&	0.9991	\\
\end{pmatrix}$ & 
$\begin{pmatrix}
-0.4524	&	0.0002	&	0.0000	\\
-0.4984	&	0.5651	&	0.7995	\\
0.7395	&	-0.8250	&	0.6006	\\
\end{pmatrix}$ \\
\\
Eigenvalues & 
$\begin{pmatrix}
-0.0029\\
-0.2479\\
-0.1925\\
\end{pmatrix}$ &
$\begin{pmatrix}
-0.0005\\
-0.8364\\
-0.5413\\
\end{pmatrix}$ \\
\\
\hline
\end{tabular}
\end{table}



\subsection{Resiliency}

The observed mean-reversion properties of the order book models may be interpreted as a form of market resiliency. After a liquidity shock, the market mechanisms work by adjusting supply so that the liquidity properties of the book tend towards an equilibrium state. We will illustrate this with impulse response analysis where one of the variables of an estimated model is perturbed from its equilibrium value and then observing how the dynamic structure of the model operates to restore the equilibrium.

As noted in the previous section, the estimated models are stationary which implies the existence of the limit $\bar\xi_\infty = \lim_{t\to\infty}E\xi_t$. Taking expectations on both sides of \eqref{eq:lin} we see that the expectation $\bar\xi_t=E\xi_t$ satisfies
\begin{equation*}
\frac{d}{dt}{\bar\xi}_t=A\bar\xi_t+a.
\end{equation*}
The equilibrium value $\bar\xi_\infty$ solves $A\xi + a=0$ which has a unique solution when the eigenvalues of $A$ are all zero. For TDC, we get $\bar\xi_\infty=(5.5069,-0.0084,-0.0229)$ and for Moeller-Maersk $\bar\xi_\infty=(10.9150,0.1148,-0.1735)$. Equilibrium properties of limit order books have been studied also e.g.\ in Luckock~\cite{luc3} and Cont et al.~\cite{cst8} but their results are not directly comparable to the above since the liquidity factors $\beta^+$ and $\beta^-$ have been deseasonalized; see Section~\ref{sec:idp}. 

\begin{figure}
\centering
\subfigure[TDC]
{
\epsfig{file=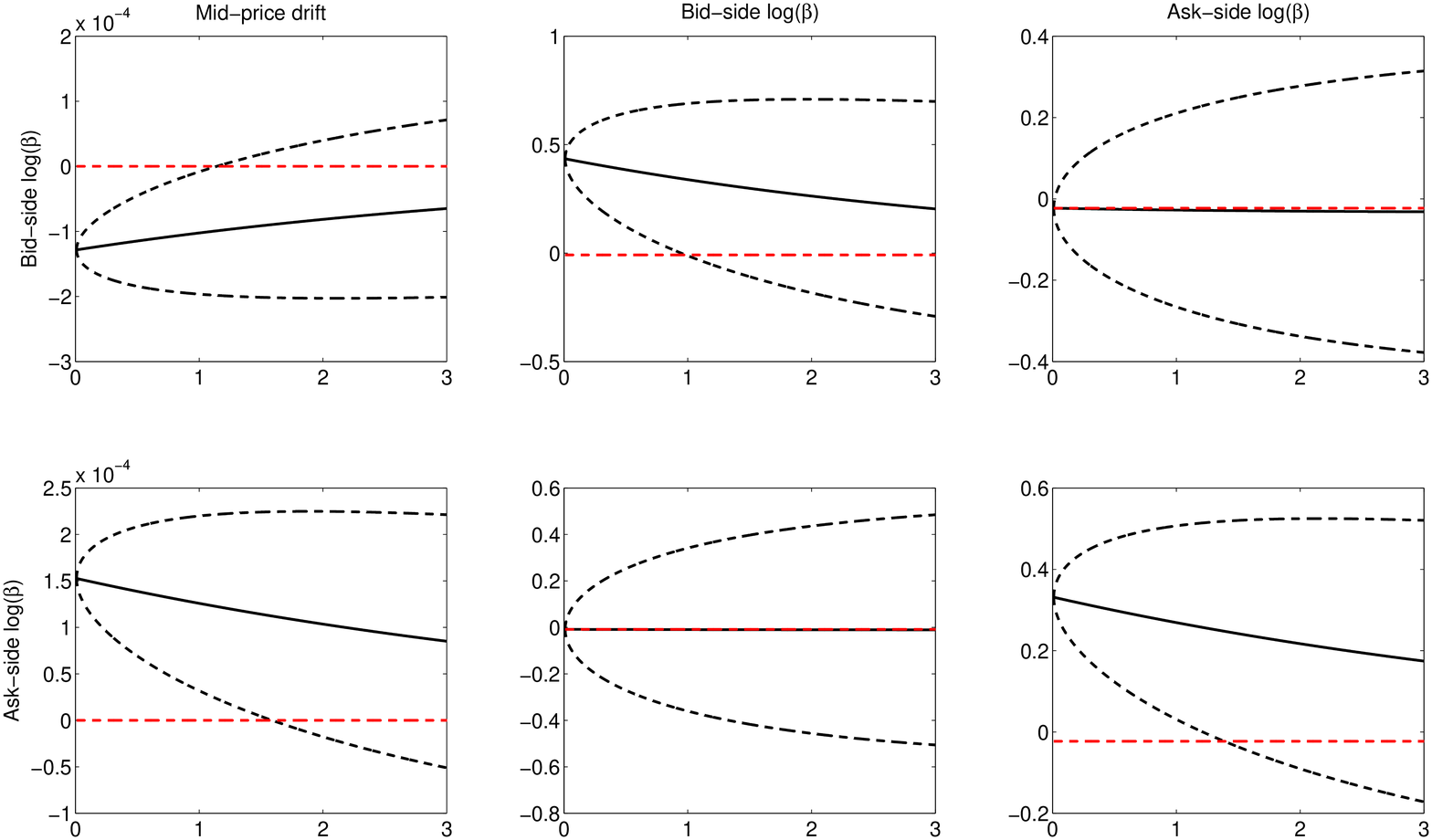,height=8cm,width=\linewidth, angle=0}
}
\subfigure[Moeller-Maersk]
{
\epsfig{file=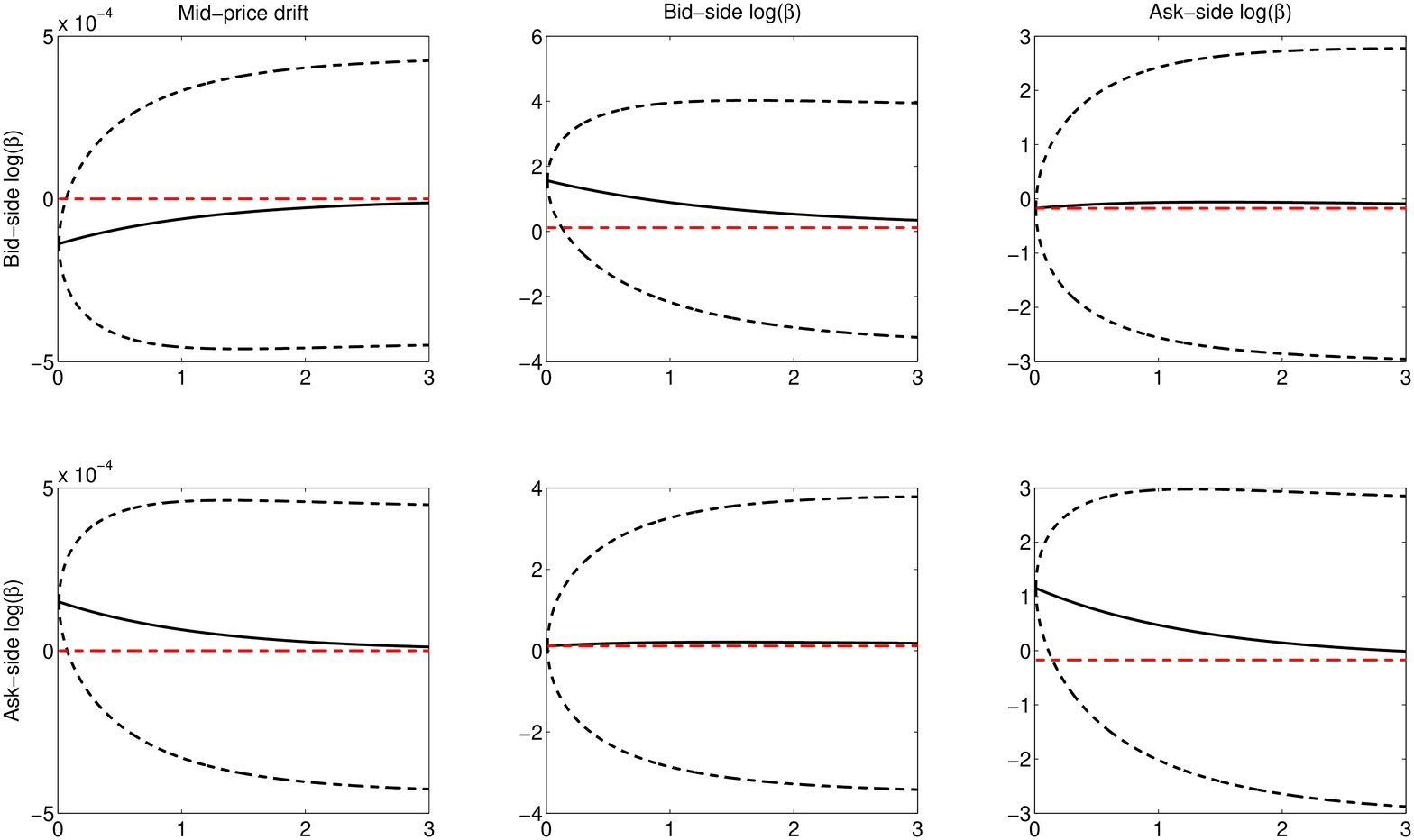,height=8cm,width=\linewidth, angle=0}
}
\caption{Impulse responses for TDC and Moeller-Maersk order book models. The first row of figures gives the responses to an impulse on the bid-side impact factor. The response variables are the mid-price drift and the logarithmic bid- and ask-side liquidity factors. The second row gives the responses to an impulse on the ask-side liquidity factor. The time unit is 10 minutes}\label{fig:resiliency}
\end{figure}

Figure~\ref{fig:resiliency} plots the impulse responses for TDC and Moeller-Maersk. In both cases we perturbed the liquidity factors $\beta_t^-$ and $\beta_t^+$ and tracked the effects on the mid-price drift and the liquidity factors. Positive deviations of the liquidity factors from their equilibrium values can be interpreted as liquidity shocks which reduce the liquidity on one side of the book. The first row of figures draws the responses to a shock on the bid-side and the second row on the ask-side. Each shock is defined as an increase of one standard deviation from the equilibrium value of the liquidity factor. The responses are given in terms of the development of the median and the 95\%-confidence interval.

A positive deviation in $\beta^-$  results in significant decreases in the subsequent drift of the  mid-price. The decreased drift can be interpreted as decreased average returns. It takes about 30 minutes for $\beta_t^-$ to return to its equilibrium value. There is practically no effect on the ask-side liquidity. The results are similar when perturbing ask-side liquidity factor $\beta_t^+$ but the effect on the mid-price is reversed. This kind of behavior of the order book is consistent with the crowding out effect which, in microstructural analysis, is usually described in terms of incoming orders.



\section{Conclusion \label{sec:conc}}

This paper proposes a reduced form approach for modeling limit order markets. The models are obtained by augmenting classical perfectly liquid market models by few additional risk factors that describe the liquidity properties of the order book. The proposed models are easy to calibrate and to analyze using standard techniques for multivariate stochastic processes. Despite the simple structure, the models are able to capture several empirically observed properties of order book dynamics. 

The modeling approach was illustrated by fitting three-dimensional continuous-time order book models to limit order data from Copenhagen stock exchange. The underlying risk factors in the models were the mid-price, and two liquidity factors describing the overall liquidity on each side of the order book. The estimation results support the chosen parameterization by revealing significant interactions between the liquidity factors and log-returns on the mid-price. The estimated model provides also a simple quantification of market resiliency with respect to liquidity shocks.

While the proposed parsimonious approach will inevitably fail to capture some details of limit order markets, we believe that its simplicity will be an advantage when studying the costs of dynamic trading strategies. The monotonicity and convexity of the cost functions in the model are essential in that respect; see e.g.\ Almgren and Chriss~\cite{ac}, Huberman and Stanzl~\cite{hs5} or Alfonsi, Schied and Fruth~\cite{asf10}. Moreover, in situations where analytic solutions are not available, the low-dimensionality of our models makes them amenable to numerical techniques such as those proposed in Bertsimas and Lo~\cite{bl98} or Koivu and Pennanen~\cite{kp10}. This is currently under investigation and will be reported elsewhere.



\bibliographystyle{plain}
\bibliography{books}

\end{document}